# 2D Potts Model Correlation Lengths: Numerical Evidence for $\xi_o = \xi_d$ at $\beta_t$

*Wolfhard Janke* and *Stefan Kappler*

Institut für Physik, Johannes Gutenberg-Universität Mainz
Staudinger Weg 7, 55099 Mainz, Germany

**Abstract**

We have studied spin-spin correlation functions in the ordered phase of the two-dimensional $q$-state Potts model with $q = 10$, 15, and 20 at the first-order transition point $\beta_t$. Through extensive Monte Carlo simulations we obtain strong numerical evidence that the correlation length in the ordered phase agrees with the exactly known and recently numerically confirmed correlation length in the disordered phase: $\xi_o(\beta_t) = \xi_d(\beta_t)$. As a byproduct we find the energy moments in the ordered phase at $\beta_t$ in very good agreement with a recent large $q$-expansion.

# 1 Introduction

First-order phase transitions have been the subject of increasing interest in recent years. They play an important role in many fields of physics as is witnessed by such diverse phenomena as ordinary melting, the quark deconfinement transition or various stages in the evolution of the early universe [1]. Even though there exists already a vast literature on this subject [2], many properties of first-order phase transitions still remain to be investigated in detail. Examples are finite-size scaling (FSS) [3], the shape of energy or magnetization distributions [4, 5], partition function zeros [6], etc., which are all closely interrelated. An important approach to attack these problems are computer simulations [7]. Here the available system sizes are necessarily limited and reliable FSS analyses are of utmost importance. The distinguishing input parameter in such analyses is the correlation length $\xi$ which sets the relevant length scale of the system and thus the range of validity of the commonly employed FSS Ansätze [3].

The well-known paradigm to investigate these questions in detail is the $q$-state Potts model [8] which can be tuned from a second-order through weakly first-order to a strong first-order transition by varying the number of states $q$. In two dimensions many quantities of interest are known exactly [9] and have been shown to have interesting algebraic interpretations in knot theory [10]. The theoretical knowledge of correlation lengths, however, is still quite limited. Only in the disordered phase exactly *at* the first-order transition point $\beta_t$ ($q \geq 5$) an explicit formula for the correlation length $\xi_d(\beta_t)$ is known [11, 12]. By analogy with the Ising model [13] and on the basis of previous numerical data [14] it has been speculated [12] that the correlation length in the ordered phase is given by $\xi_o(\beta_t) = \frac{1}{2}\xi_d(\beta_t)$. In fact, very recently this could be proven for a certain definition $\xi_{o,1}$, but the possibility of a larger correlation length $\xi_{o,2} \geq \xi_{o,1}$ was left open [15]. In this note we report a careful Monte Carlo study of the leading correlation length in the two phases which yields unambiguous numerical evidence *against* this conjecture. Rather, our numerical analysis strongly favors an earlier conjecture that $\xi_o(\beta_t) = \xi_d(\beta_t)$.

## 2 Model and simulation

The Potts model is defined by the partition function

$$Z = \sum_{\{s_i\}} e^{-\beta E};\ E = -\sum_{\langle ij \rangle} \delta_{s_i s_j};\ s_i = 1, \ldots, q, \tag{1}$$

where $i$ denote the lattice sites, $\langle ij \rangle$ are nearest-neighbor pairs, and $\delta_{s_i s_j}$ is the Kronecker delta symbol. We simulated the model with $q = 10$, 15, and 20 in the ordered phase at $\beta_t = \ln(1 + \sqrt{q})$ on 2D lattices of size $V = L_x \times L_y$. For each $q$ we considered two different lattice geometries, namely $L \times L$ and $2L \times L$ lattices with $L = 150$, 60, and 40 for $q = 10$, 15, and 20, respectively. To take advantage of translational invariance we used periodic boundary conditions. From our experience with simulations in the disordered phase we knew that lattice sizes $L \approx 14\xi_d$ are large enough to suppress tunneling events. In fact, starting from a completely ordered configuration, we never observed a tunneling event into the disordered phase, thus allowing statistically meaningful pure-phase measurements of energy moments and correlation functions. In Fig. 1 we show the probability distributions $P(e)$ of the energy (normalized to unit area) in the ordered as well as in the disordered phase, demonstrating that the two peaks are indeed very well separated.

While in our earlier study [16] of the disordered phase we employed the single-cluster algorithm, here we found it more efficient, in terms of real-time performance on a Cray-YMP, to update the spins with a vectorized standard heat-bath algorithm. With this update algorithm the value of the spins in the largest (spanning) cluster never changed, as would clearly be the case with fixed boundary conditions. The main difference between the two types of boundary conditions is that, with fixed boundary spins, excitations of disordered bubbles would be repelled from the walls, while with periodic boundary conditions they can freely move around the torus. The statistic parameters of the runs are collected in Table 1, where we give the number of update sweeps in units of the integrated autocorrelation time of the energy, $\tau_{\mathrm{int,e}}$, which is roughly the same for the two lattice geometries.

To determine the correlation length $\xi_o$ we followed Gupta and Irbäck [14] and considered the $k_y^{(n)} = 2\pi n / L_y$ momentum projections ($i = (i_x, i_y)$),

$$g^{(n)}(i_x, j_x) = \frac{1}{L_y} \sum_{i_y, j_y} \langle \delta_{s_i s_j} - \frac{1}{q} \rangle e^{i k_y^{(n)} (i_y - j_y)}. \tag{2}$$

For $n \neq 0$ the projections are free of constant background terms and, similar to our analysis in the disordered phase, we tried to determine $\xi_o$ from fits of $g^{(n)}(x) \equiv g^{(n)}(i_x, 0)$ to the Ansatz

$$g^{(n)}(x) = a\,\mathrm{ch}(\frac{L_x/2 - x}{\xi_o^{(n)}}) + b\,\mathrm{ch}(c\frac{L_x/2 - x}{\xi_o^{(n)}}), \tag{3}$$

with $\xi_o^{(n)} = \xi_o/\sqrt{1+(2\pi n \xi_o/L_y)^2}$. Tests for the Ising model suggested that this is a good approximation for $L/n\xi > 10$. For the Potts models with $L/\xi \approx 14$ we expect that $\xi_o^{(1)}$ is smaller than $\xi_o$ by about 10% for the $k_y$-projections and 2% for the $k_x$-projection on the $2L \times L$ lattice, respectively. Even though in the ordered phase it is more efficient to update the spins with a heat-bath algorithm, for the measurements we decomposed the spin configuration into stochastic Swendsen-Wang cluster and used the cluster estimator $\langle \delta_{s_i s_j} - 1/q \rangle = (1 - 1/q)\langle \Theta(i,j) \rangle$, where $\Theta(i,j) = 1$, if $i$ and $j$ belong to the same cluster, and $\Theta = 0$ otherwise. All error bars are estimated by means of the jack-knife technique.

# 3 Results

We first checked that the average energy agrees with the exact result [9], and compared the second and third energy moments, $c_o = \beta_t^2 \mu_o^{(2)} = \beta_t^2 V \langle (e - \langle e \rangle)^2 \rangle$ and $\mu_o^{(3)} = V^2 \langle (e - \langle e \rangle)^3 \rangle$, with the large $q$ expansions of Ref.[5]; cp. Table 2. Even for our smallest value $q = 10$ and the third moment we observe a very good agreement with the (Padé resummed) large $q$ expansion. In addition we looked at the magnetization $m = (q\,\langle \max\{n_i\} \rangle / V - 1)/(q-1)$ and its cluster estimator $m' = \langle |C|_{\mathrm{span}} \rangle / V$, where $n_i$ denotes the number of spins of "orientation" $i = 1, \ldots, q$ and $|C|_{\mathrm{span}}$ is the size of the largest (spanning) cluster. Table 3 shows that the two estimators give almost identical results which agree very well with the exact expression $m = \prod_{n=1}^{\infty} [(1 - x^n)/(1 - x^{4n})]$, where, for $q \geq 5$, $x$ is defined by $q = x + 2 + x^{-1}$ $(0 < x < 1)$ [17]. Also shown is the susceptibility[1], $\chi_o = m_o^{(2)} = V \langle (m - \langle m \rangle)^2 \rangle$, and the third moment, $m_o^{(3)} = V^2 \langle (m - \langle m \rangle)^3 \rangle$.

[1] To conform with the present normalization, our Monte Carlo estimate for $\chi_d$ [16] should be multiplied by a factor of $1/(q-1)$.

For the estimate of $\xi_o(\beta_t)$ we concentrated on the $k_y = 2\pi/L_y$ projection $g^{(1)}(x)$. The qualitative behaviour of $g^{(1)}$ is illustrated in the semi-log plots of Fig. 2. For comparison also our previous results [16] for $g^{(0)}$ in the disordered phase are shown. Already these plots suggest that the two correlation functions are governed by the same asymptotic decay law, i.e., that $\xi_o(\beta_t) = \xi_d(\beta_t)$. In fact, the dotted lines interpolating the $g^{(1)}$ data are constrained fits to the Ansatz (3) assuming that $\xi_o = \xi_d$ ($= 10.559519\ldots$, $4.180954\ldots$, $2.695502\ldots$ for $q = 10$, 15, 20 [11, 12]). To be sure we also performed unconstrained four-parameter fits using Ansatz (3). Figure 2 shows that, despite our high statistics, in the ordered phase it was impossible to get reliable estimates of $g^{(1)}$ for very large distances. Compared with our analysis in the disordered phase the available fit intervals are consequently shifted to smaller $x$. This makes the fits in the ordered phase somewhat more sensitive to higher-order excitations and, based on our previous experience with varying fit intervals, we estimate that the numbers for $\xi_o^{(1)}$ collected in Table 4 should underestimate the true value by about $15 - 30\%$. For a comparison of $\xi_o(\beta_t)$ with $\xi_d(\beta_t)$ we have therefore used approximately the same fit intervals. For example, for $q = 10$, $L \times L$ lattice and the fit starting at $x_{\rm min} = 11$, we obtain from Table 4 $\xi_o(\beta_t) = 7.9(9)$. Recalling our estimate of $\xi_d(\beta_t) = 8.8(3)$ for a comparable fit in the disordered phase [16], this yields a ratio of $\xi_o(\beta_t)/\xi_d(\beta_t) = 0.9(1)$, suggesting again that $\xi_o(\beta_t) = \xi_d(\beta_t)$. This is corroborated by the corresponding estimates for $q = 15$ and $q = 20$, which are actually closer to unity.

To make this statement even more convincing we have plotted in Fig. 3 the ratio $\xi_o^{\rm eff}/\xi_d^{\rm eff}$, where $\xi^{\rm eff} = 1/\ln[g(x)/g(x+1)]$ is the usual effective correlation length (with the correction $\xi_o = \xi_o(\xi_o^{(1)})$ already taken into account). Here we have again implicitly assumed that higher-order excitations play a similar role in both phases and effectively drop out when plotting the ratio of the two correlation lengths. Figure 3 shows that, over a sizeable range of distances, $\xi_o^{\rm eff}/\xi_d^{\rm eff} = 1$, with an uncertainty of only about 5%. We interpret this as further support of our conjecture that $\xi_o(\beta_t) = \xi_d(\beta_t)$.

# 4 Discussion

To summarize, applying a non-zero momentum projection to the correlation function in the ordered phase we determined $\xi_o(\beta_t)$ with an accuracy compa-rable to previous investigations in the disordered phase [16]. By comparing the two correlation lengths at $\beta_t$ we obtain strong numerical evidence that $\xi_o = \xi_d$. At first sight this is in striking disagreement with a very recent exact proof [15] of the earlier conjecture $\xi_o = \frac{1}{2}\xi_d$ for one definition of the ordered correlation length, $\xi_{o,1}$. For another definition, $\xi_{o,2}$, however, only the relation $\xi_{o,2} \geq \xi_{o,1}$ could be established in Ref.[15]. This is clearly consistent with our result if we identify the numerically determined $\xi_o$ with $\xi_{o,2}$. We are currently investigating this problem in more detail [18] by using precisely the definitions of Ref.[15] which are based on geometrical properties of Potts model clusters such as, e.g., the distribution function of the cluster diameter.

## Acknowledgements

WJ thanks the DFG for a Heisenberg fellowship and SK gratefully acknowledges a fellowship by the Graduiertenkolleg "Physik and Chemie supramolekularer Systeme". Work supported by computer grants hkf001 of HLRZ Jülich and bvpf03 of Norddeutscher Vektorrechnerverbund (NVV) Berlin-Hannover-Kiel.

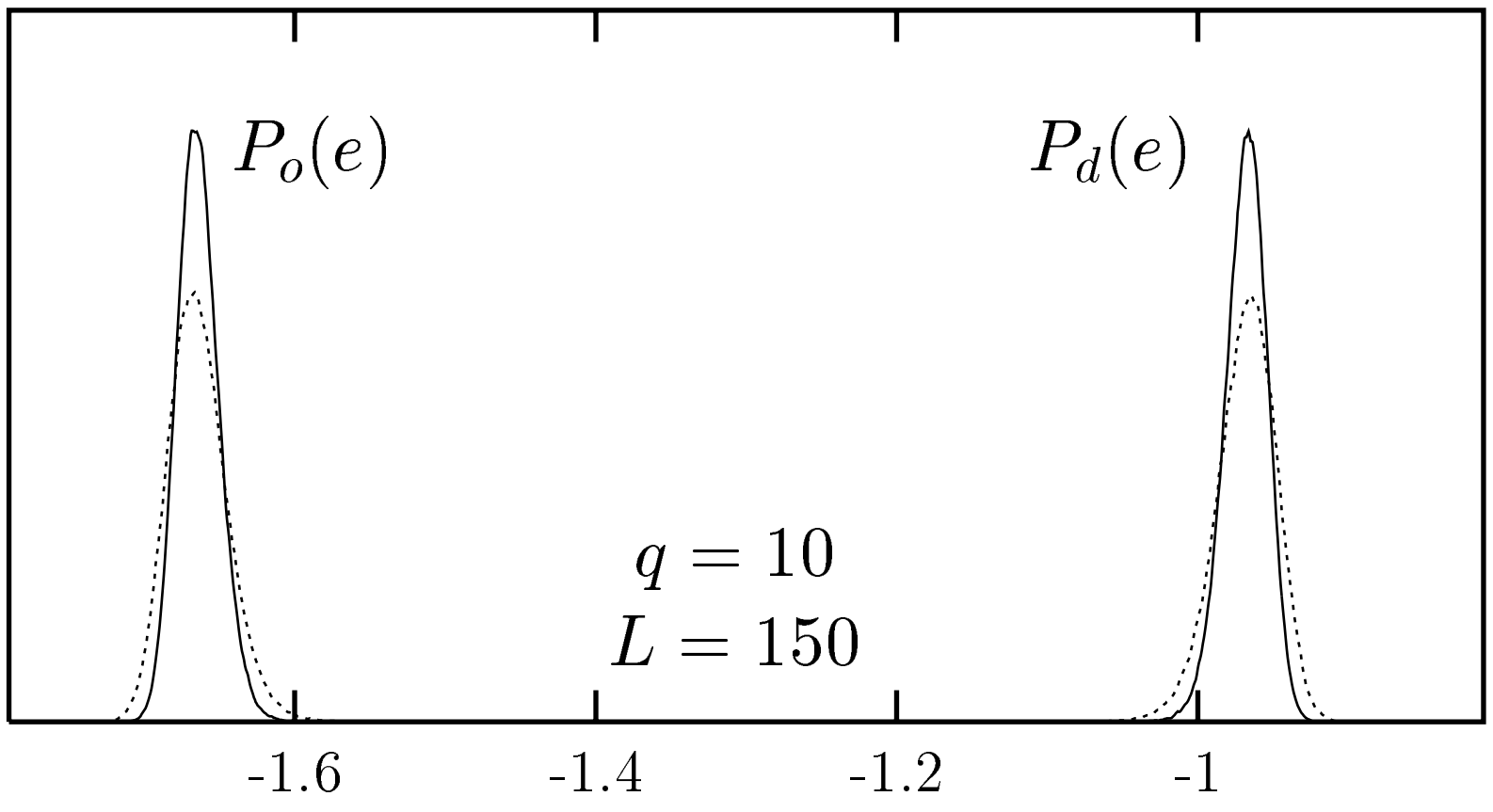

Figure 1: Probability distribution $P(e)$ in the two phases for $L \times L$ (dotted line) and $2L \times L$ lattices.

Table 1: Integrated autocorrelation time $\tau_{\mathrm{int,e}}$ of the energy and the number of update sweeps in units of $\tau_{\mathrm{int,e}}$.

| | $q = 10$ | $q = 15$ | $q = 20$ |
|---|---|---|---|
| $\tau_{\mathrm{int,e}}$ | $\approx 140$ | $\approx 20$ | $\approx 10$ |
| $L \times L$ | 118 000 | 1 280 000 | 1 280 000 |
| $2L \times L$ | 121 000 | 1 280 000 | 5 120 000 |

Table 2: Comparison of numerical and analytical results for energy moments at $\beta_t$ in the ordered phase.

| Observable | $q = 10$ | $q = 15$ | $q = 20$ |
|---|---|---|---|
| $e_o$ (MC, $L \times L$) | $-1.664177(81)$ | $-1.765850(34)$ | $-1.820722(43)$ |
| $e_o$ (MC, $2L \times L$) | $-1.664262(57)$ | $-1.765875(26)$ | $-1.820689(14)$ |
| $e_o$ (exact) | $-1.664253...$ | $-1.765906...$ | $-1.820684...$ |
| $c_o$ (MC, $L \times L$) | 17.95(13) | 8.016(21) | 5.351(15) |
| $c_o$ (MC, $2L \times L$) | 17.81(10) | 8.004(19) | 5.3612(55) |
| $c_o$ (large $q$) | 18.1(1) | 8.00(3) | 5.362(3) |
| $\mu_o^{(3)}$ (MC, $L \times L$) | 1979(87) | 180.5(3.1) | 57.0(1.3) |
| $\mu_o^{(3)}$ (MC, $2L \times L$) | 1836(71) | 189.7(5.1) | 56.24(40) |
| $\mu_o^{(3)}$ (large $q$) | 1850(40) | 179(4) | 56.8(4) |

Table 3: Comparison of numerical and analytical results for magnetization moments at $\beta_t$ in the ordered phase.

| Observable | $q = 10$ | $q = 15$ | $q = 20$ |
|---|---|---|---|
| $m$ (MC, $L \times L$) | 0.857047(71) | 0.916631(21) | 0.941199(21) |
| $m'$ (MC, $L \times L$) | 0.857047(71) | 0.916634(21) | 0.941197(21) |
| $m$ (MC, $2L \times L$) | 0.857113(49) | 0.916648(16) | 0.9411782(66) |
| $m'$ (MC, $2L \times L$) | 0.857113(49) | 0.916648(16) | 0.9411791(66) |
| $m$ (exact) | 0.857106... | 0.916663... | 0.9411759... |
| $\chi_o$ (MC, $L \times L$) | 4.750(60) | 0.8090(36) | 0.3348(17) |
| $\chi_o$ (MC, $2L \times L$) | 4.663(43) | 0.8095(38) | 0.33509(55) |
| $m_o^{(3)}$ (MC, $L \times L$) | −1521(85) | −45.9(1.2) | −8.55(32) |
| $m_o^{(3)}$ (MC, $2L \times L$) | −1372(62) | −49.4(2.2) | −8.321(88) |

Table 4: Numerical estimates of the correlation length $\xi_o^{(1)}(\beta_t)$ from four-parameter fits to the Ansatz (3) in the range $x_{\min} - x_{\max}$. For the $2L \times L$ lattices the fits along the $x$ and $y$ direction are distinguished by the index.

| | $q = 10$, $L = 150$ | | $q = 15$, $L = 60$ | | $q = 20$, $L = 40$ | |
|---|---|---|---|---|---|---|
| Lattice | $x_{\min} - x_{\max}$ | $\xi_o^{(1)}$ | $x_{\min} - x_{\max}$ | $\xi_o^{(1)}$ | $x_{\min} - x_{\max}$ | $\xi_o^{(1)}$ |
| $L \times L$ | 4−54 | 7.4(4) | 3−19 | 3.0(1) | 2−12 | 2.0(1) |
| | 7−54 | 7.5(6) | 5−19 | 3.2(3) | 3−12 | 2.0(2) |
| | 11−54 | 7.5(8) | 7−19 | 3.4(9) | 4−12 | 2.0(2) |
| $(2L \times L)_x$ | 4−40 | 7.1(3) | 3−22 | 3.2(1) | 2−14 | 2.0(1) |
| | 7−40 | 7.4(4) | 5−22 | 3.2(2) | 3−14 | 2.0(1) |
| | 11−40 | 7.3(5) | 7−22 | 3.1(2) | 4−14 | 2.0(2) |
| $(2L \times L)_y$ | 4−40 | 7.4(3) | 3−22 | 3.2(2) | 2−14 | 2.1(1) |
| | 7−40 | 7.5(4) | 5−22 | 3.2(2) | 3−14 | 2.2(1) |
| | 11−40 | 7.4(5) | 7−22 | 3.2(4) | 4−14 | 2.2(2) |

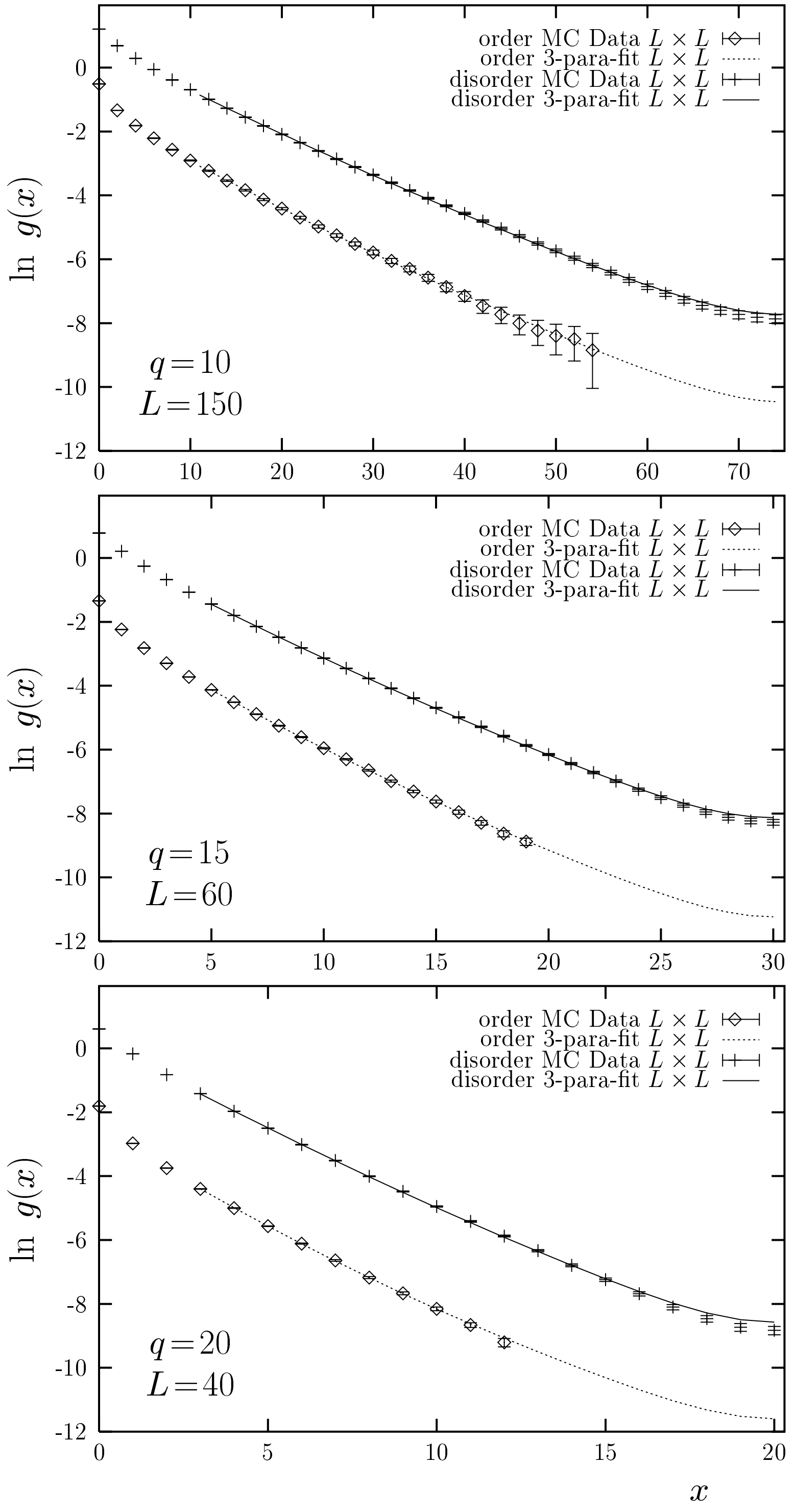

Figure 2: Semi-logarithmic plots of the projected correlation function $g^{(1)}$ for $q = 10$, 15 and 20 in the ordered phase. For comparison also $g^{(0)}$ in the disordered phase [16] is shown.

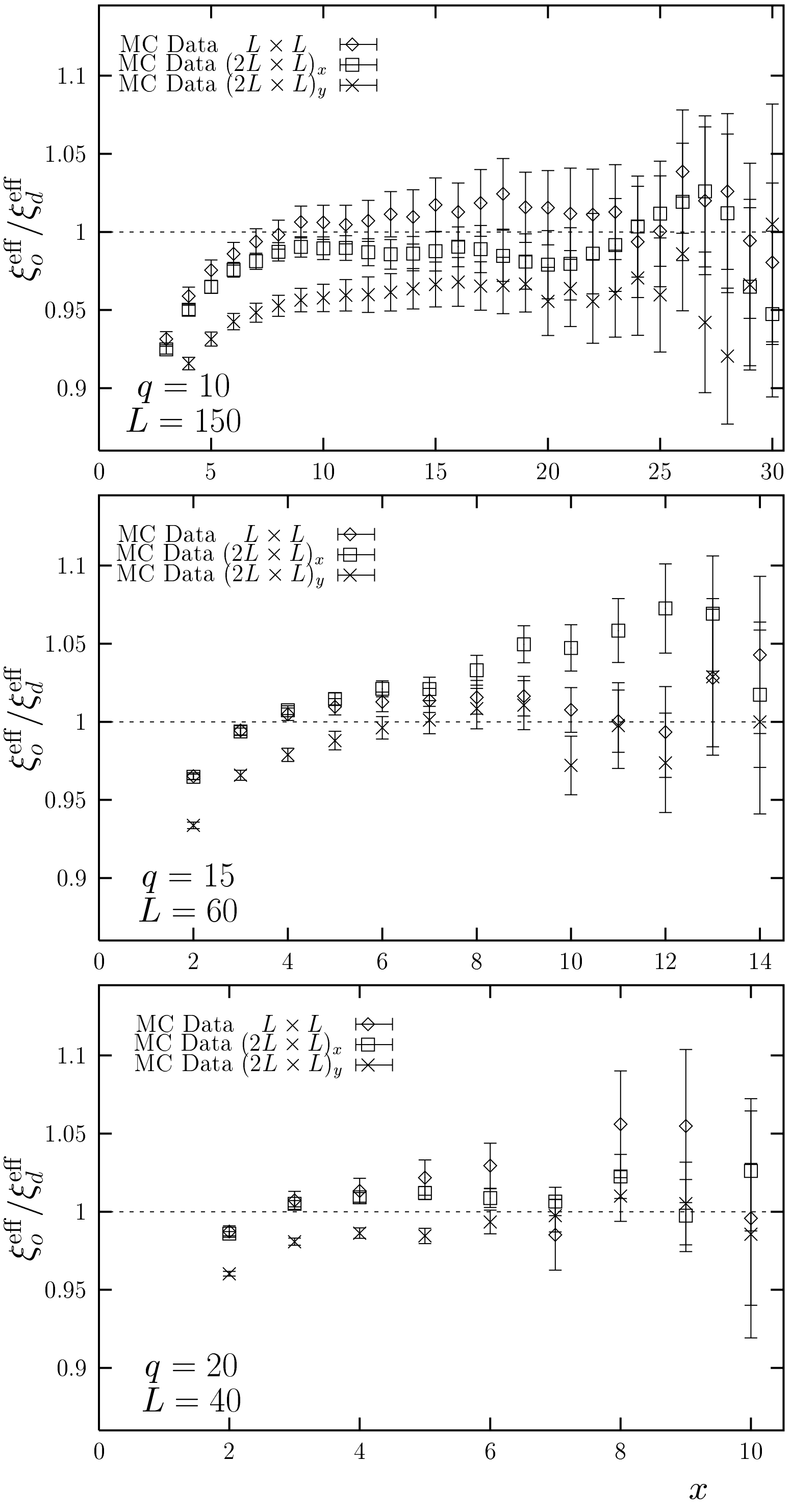

Figure 3: Ratio of effective correlation lengths in the ordered and disordered phase for $q = 10$, 15 and 20.